\documentclass[prd,showpacs,showkeys,floatfix,amsmath,amssymb,floatfix,english]{revtex4}

\usepackage{amsfonts}
\usepackage{subfig}
\usepackage{dcolumn}
\usepackage{natbib}
\usepackage{bm}
\usepackage{booktabs}
\usepackage{slashed}

\usepackage[T1]{fontenc}
\usepackage[utf8]{inputenc}
\usepackage{babel}
\usepackage{units}
\usepackage{amsmath}
\usepackage{esint}
\usepackage{multirow}
\usepackage{graphicx}
\usepackage[normalem]{ulem}
\usepackage[colorlinks, citecolor=blue,anchorcolor=red,menucolor=red, linkcolor=red,filecolor=red,runcolor=red,urlcolor=blue,frenchlinks=red]{hyperref}

\usepackage{ulem}
\usepackage{color}

\newcommand{\chadd}[1]{\textcolor{blue}{#1}}

\makeatletter
\newcommand{\figcaption}{\def\@captype{figure}\caption}
\newcommand{\tabcaption}{\def\@captype{table}\caption}

\newcommand{\Rmnum}[1]{\expandafter\@slowromancap\romannumeral #1@}
\def\hlinewd#1{%
  \noalign{\ifnum0=`}\fi\hrule \@height #1 \futurelet
   \reserved@a\@xhline}
\makeatother
\begin{document}

\title{Triply heavy $QQ\bar Q\bar q$ tetraquark states}

\author{Jin-Feng Jiang}
%\email{jfjiang@pku.edu.cn}
\affiliation {School of Physics and State Key Laboratory of Nuclear
Physics and Technology, Peking University, Beijing 100871, China}

\author{Wei Chen}
\email{chenwei29@mail.sysu.edu.cn} \affiliation{School of Physics,
Sun Yat-Sen University, Guangzhou 510275, China}

\author{Shi-Lin Zhu}
\email{zhusl@pku.edu.cn} \affiliation{
School of Physics and State Key Laboratory of Nuclear Physics and Technology, Peking University, Beijing 100871, China \\
Collaborative Innovation Center of Quantum Matter, Beijing 100871, China \\
Center of High Energy Physics, Peking University, Beijing 100871,
China}

\begin{abstract}
Within the framework of QCD sum rules, we have investigated the
tetraquark states with three heavy quarks. We systematically
construct the interpolating currents for the possible
$cc\bar{c}\bar{q}$, $cc\bar{b}\bar{q}$, $bc\bar{b}\bar{q}$,
$bb\bar{b}\bar{q}$ tetraquark states with quantum numbers
$J^{P}=0^{+}$ and $J^{P}=1^{+}$. Using these interpolating currents,
we have calculated the two-point correlation functions and extracted
the mass spectra for the above tetraquark states. We also discuss
the decay patterns of these tetraquarks, and notice that the
$cc\bar{c}\bar{q}$, $cc\bar{b}\bar{q}$, $bc\bar{b}\bar{q}$ may decay
quickly with a narrow width due to their mass spectra. The
$bb\bar{b}\bar{q}$ tetraquarks are expected to be very narrow
resonances since their OZI-allowed decay modes are kinematically
forbidden. These states may be searched for in the final states with
a $B$ meson plus a light meson or photon.
\end{abstract}

 \maketitle

%%%%%%%%%%%%%%%%%%%%%%%%%%%%%%%%%%%%%%%
\section{INTRODUCTION}
\label{sec2}
%%%%%%%%%%%%%%%%%%%%%%%%%%%%%%%%%%

The existence of multiquark states was first suggested at the birth
of the quark model~\cite{1964-Gell-Mann-p214-215,1964-Zweig-p-}. In
the past fifty years, the search of the multiquark matter has been
an extremely intriguing research issue, which shall provide
important hints to the understanding of the non-perturbative
QCD~\cite{2016-Chen-p1-121,
2006-Swanson-p243-305,2008-Voloshin-p455-511,2016-Chen-p406-421,2016-Esposito-p1-97,
2017-Lebed-p143-194,2017-Guo-p-}.

Tetraquarks are composed of diquarks and antidiquarks. They are
compact hadron states bound by colored force between four quarks.
The light tetraquark configurations $q\bar q q \bar q$ and $q\bar q
s\bar s$ were proposed to explain the scalar mesons below 1
GeV~\cite{1977-Jaffe-p281-281,2007-Chen-p94025-94025,
1999-Black-p74026-74026,2004-Maiani-p212002-212002}. In the heavy
sector, the exotic charmed mesons
$D_{s0}^\ast(2317)$~\cite{2003-Aubert-p242001-242001}  and
$D_{s1}(2460)$~\cite{2003-Besson-p32002-32002} were studied as the
singly charmed-strange tetraquark states in
Refs.~\cite{2003-Cheng-p193-200,2004-Dmitrasinovic-p96011-96011,2005-Maiani-p14028-14028}.
Very recently, the D\O~Collaboration reported a narrow structure in
the $B_s^0\pi^\pm$ invariant mass
spectrum~\cite{2016-Abazov-p22003-22003}. This charged $X(5568)$
meson, if it really exists, could be a good candidate for a
tetraquark state consisting of four different flavor quarks
$su\bar{d}\bar{b}$ (or $sd\bar{u}\bar{b}$). Various of theoretical
configurations have been proposed to investigate the nature of
$X(5568)$, such as the diquark-antidiquark tetraquark
state~\cite{2016-Chen-p22002-22002,2016-Agaev-p74024-74024,2016-Wang-p335-339,2016-Burns-p627-633,
2016-Guo-p593-595}, hadron
molecule~\cite{2016-Agaev-p351-351,2016-Kang-p54010-54010}, and so
on. The detailed introduction of this state can be found in the
recent review paper~\cite{2017-Chen-p76201-76201}.

The hidden flavor tetraquarks are the most extensively studied ones
in the past decade, due to the plenty of the charmonium-like and
bottomonium-like states observed in
experiments~\cite{2016-Chen-p1-121,
2016-Esposito-p1-97,2017-Lebed-p143-194,2017-Guo-p-,2017-Ali-p-}.
Most of these XYZ states do not fit into the quark model spectrum
easily and their decay final products contain a heavy
quark-antiquark pair. Especially for the charged $Z_c$ ($Z_b$)
states, they contain a $c\bar c$ ($b\bar b$) pair plus at least a
pair of light quark-antiquark. They were considered as very good
tetraquark candidates with two heavy quarks~\cite{2016-Chen-p1-121}.
However, some of these XYZ mesons can also be interpreted as loosely
bound molecular states, especially for those near-threshold
resonances~\cite{2016-Chen-p1-121,2017-Guo-p-}. Sometimes, it is not
so easy to distinguish a loosely hadron molecule from a
compact tetraquark state.

The doubly hidden flavor $QQ\bar Q\bar Q$ system consists of four
heavy quarks. Such a four-quark system favors the compact tetraquark
configuration than the hadron molecule because the binding force
comes from the short-range gluon exchange and the light mesons can
not be exchanged between two quarkonium states. Recently, several
collaborations reported the observations of the $J/\psi$ pairs at
LHCb~\cite{2012-Aaij-p52-59}, D\O~\cite{2014-Abazov-p111101-111101}
and CMS~\cite{2014-Khachatryan-p94-94}, $\Upsilon(1S)$ pair at
CMS~\cite{2017-Khachatryan-p13-13} and the simultaneous
$J/\psi\Upsilon(1S)$ events at D\O~\cite{2016-Abazov-p82002-82002}
and CMS~\cite{CMS1}. These observations have trigged theoretical
discussions of the $QQ\bar Q\bar Q$ tetraquark
systems~\cite{2016-Chen-p-,2016-Wu-p-,2017-Karliner-p34011-34011,2016-Bai-p-,2017-Wang-p-a}.
It is natural to investigate the existence of the triply heavy
$QQ\bar Q\bar q$ tetraquark states~\cite{2017-Chen-p5-5} in the same
schemes. Actually, the associated production of bottomonia and open
charm hadrons $\Upsilon D/D_s$ in $pp$ collisions was recently
reported by the LHCb Collaboration~\cite{2016-Aaij-p52-52}. In this
work, we shall study the mass spectra of the possible triply heavy
$QQ\bar Q\bar q$ tetraquark states in the QCD sum rule method.

This paper is organized as follows. In Sect.~\Rmnum{2}, we construct
the triply heavy tetraquark interpolating currents and introduce the
formalism of QCD sum rules. The spectral densities for these
currents are then evaluated and collected in
Appendix~\ref{spectral_fucntion}. In Sect.~\Rmnum{3}, we perform
numerical analyses for the tetraquark mass sum rules and extract the
mass spectra for the $cc\bar{c}\bar{q}$, $cc\bar{b}\bar{q}$,
$bc\bar{b}\bar{q}$, $bb\bar{b}\bar{q}$ systems. We also discuss the
possible decay patterns of these tetraquark states. The last section
is a brief summary.

%%%%%%%%%%%%%%%%%%%%%%%%%%%%%%%%%
\section{Formalism}
\label{sec2}
%%%%%%%%%%%%%%%%%%%%%%%%%%%%%%%%%%

The starting point of the QCD sum
rule~\cite{1979-Shifman-p385-447,1985-Reinders-p1-1,2000-Colangelo-p1495-1576}
is the interpolating current which couples to the hadrons with the
same quantum numbers. We construct the interpolating currents for
the triply heavy tetraquarks with quantum numbers $J^{P}=0^{+}$,
\begin{align}
J_{1} & =Q_{1a}^{T}C\gamma_{5}Q_{2b}\left(\bar{Q}_{3a}\gamma_{5}C\bar{q}_{b}^{T}+\bar{Q}_{3b}\gamma_{5}C\bar{q}_{a}^{T}\right),\nonumber \\
J_{2} & =Q_{1a}^{T}C\gamma_{\mu}Q_{2b}\left(\bar{Q}_{3a}\gamma^{\mu}C\bar{q}_{b}^{T}+\bar{Q}_{3b}\gamma^{\mu}C\bar{q}_{a}^{T}\right), \label{currents0+}
\nonumber \\
J_{3} & =Q_{1a}^{T}C\gamma_{5}Q_{2b}\left(\bar{Q}_{3a}\gamma_{5}C\bar{q}_{b}^{T}-\bar{Q}_{3b}\gamma_{5}C\bar{q}_{a}^{T}\right),\nonumber \\
J_{4} &
=Q_{1a}^{T}C\gamma_{\mu}Q_{2b}\left(\bar{Q}_{3a}\gamma^{\mu}C\bar{q}_{b}^{T}-\bar{Q}_{3b}\gamma^{\mu}C\bar{q}_{a}^{T}\right). 
\end{align}
where $Q$ denotes the heavy quark $c$ or $b$ and $q$ denotes the
light quark $u$ or $d$. The currents $J_{1}$ and $J_{2}$ are color
symmetric
$\left[\boldsymbol{6}_{c}\right]_{Q_{1}Q_{2}}\otimes\left[\bar{\boldsymbol{6}}_{c}\right]_{\bar
Q_{3}\bar q}$, and the currents $J_{3}$ and $J_{4}$ are color
antisymmetric
$\left[\bar{\boldsymbol{3}}_{c}\right]_{Q_{1}Q_{2}}\otimes\left[\boldsymbol{3}_{c}\right]_{\bar
Q_{3}\bar q}$.

Note that the currents $J_{2}$ and $J_{3}$ vanish in the $QQ\bar
Q^{\prime}\bar q$ systems when the two heavy quarks in the diquark
$QQ$ have the same flavors ($cc$ or $bb$), due to Fermi statistics~\cite{2013-Du-p14003-14003,2014-Chen-p54037-54037}.

The interpolating currents of the tetraquarks with quantum numbers
$J^{P}=1^{+}$ are
\begin{align}
J_{1\mu} & =Q_{1a}^{T}C\gamma_{5}Q_{2b}\left(\bar{Q}_{3a}\gamma_{\mu}C\bar{q}_{b}^{T}+\bar{Q}_{3b}\gamma_{\mu}C\bar{q}_{a}^{T}\right),\nonumber \\
J_{2\mu} & =Q_{1a}^{T}C\gamma_{\mu}Q_{2b}\left(\bar{Q}_{3a}\gamma_{5}C\bar{q}_{b}^{T}+\bar{Q}_{3b}\gamma_{5}C\bar{q}_{a}^{T}\right),\nonumber \\
J_{3\mu} & =Q_{1a}^{T}C\gamma_{5}Q_{2b}\left(\bar{Q}_{3a}\gamma_{\mu}C\bar{q}_{b}^{T}-\bar{Q}_{3b}\gamma_{\mu}C\bar{q}_{a}^{T}\right),\nonumber \\
J_{4\mu} &
=Q_{1a}^{T}C\gamma_{\mu}Q_{2b}\left(\bar{Q}_{3a}\gamma_{5}C\bar{q}_{b}^{T}-\bar{Q}_{3b}\gamma_{5}C\bar{q}_{a}^{T}\right).
\end{align}
The currents $J_{2\mu}$ and $J_{3\mu}$ also vanish in the $QQ\bar
Q^{\prime}\bar q$ systems.

In this work, we shall discuss the $cc\bar{c}\bar{q}$,
$cc\bar{b}\bar{q}$, $bc\bar{b}\bar{q}$, $bb\bar{b}\bar{q}$
tetraquark systems in the framework of the QCD sum rules. The
correlation function for the scalar currents reads
\begin{equation}
\Pi\left(q\right)=i\int d^{4}x\,e^{iq\cdot x}\left\langle
0\left|T\left[j\left(x\right)j^{\dagger}\left(0\right)\right]\right|0\right\rangle\,
,
\end{equation}
and that for the vector currents
\begin{align}
\Pi_{\mu\nu}\left(q\right) & =i\int d^{4}x\,e^{iq\cdot x}\left\langle 0\left|T\left[j_{\mu}\left(x\right)j_{\nu}^{\dagger}\left(0\right)\right]\right|0\right\rangle \nonumber \\
 & =-\left(g_{\mu\nu}-\frac{q_{\mu}q_{\nu}}{q^{2}}\right)\Pi_{1}\left(q^{2}\right)+\frac{q_{\mu}q_{\nu}}{q^{2}}\Pi_{0}\left(q^{2}\right)
\end{align}
where $\Pi_{1}\left(q^{2}\right)$ and $\Pi_{0}\left(q^{2}\right)$
are the vector and scalar current polarization functions
respectively. They can be extracted by the projection,
\begin{align}
 & \Pi_{1}\left(q^{2}\right)=-\frac{1}{3}\left(g^{\mu\nu}-\frac{q^{\mu}q^{\nu}}{q^{2}}\right)\Pi_{\mu\nu}\left(q\right)\, ,\\
 & \Pi_{0}\left(q^{2}\right)=\frac{q^{\mu}q^{\nu}}{q^{2}}\Pi_{\mu\nu}\left(q\right)\, .
\end{align}

At the hadron level, the correlation function can be expressed with
the spectral function through the dispersion relation
\begin{equation}
\Pi\left(q^{2}\right)=\left(q^{2}\right)^{N}\int_{0}^{\infty}\mathrm{d}s\frac{\rho\left(s\right)}{s^{N}\left(s-q^{2}-\mathrm{i}\epsilon\right)}+\sum_{k=0}^{N-1}b_{n}\left(q^{2}\right)^{k}\,
, \label{eq:dispersion relation}
\end{equation}
where $\rho\left(s\right)$ is the spectral representation which is
related to the imaginary part of the correlation function
$\rho\left(s\right)\equiv\nicefrac{\mathrm{Im}\Pi\left(s\right)}{\pi}$.
In the QCD sum rule framework, $\rho\left(s\right)$ is always
assumed to be
\begin{align}
\rho\left(s\right) & =f_{X}^{2}\delta\left(s-m_{X}^{2}\right)\left\langle 0\left|\eta\right|n\right\rangle \left\langle n\left|\eta^{\dagger}\right|0\right\rangle \nonumber \\
 & =f_{X}^{2}\delta\left(s-m_{X}^{2}\right)+\text{continuum}\, ,
\end{align}
where $m_{X}$ is the mass of the lowest-lying resonance $X$ and
$f_{X}$ is the coupling constant.

At the QCD level, the correlation function can be calculated via the
operator product expansion (OPE)
method~\cite{1969-Wilson-p1499-1512}. To get the Wilson coefficients
we will use the propagator for the light quarks
\begin{align}
\mathrm{i}S^{q}\left(x\right) & =\frac{\mathrm{i}\delta_{ab}x\!\!\!/}{2\pi^{2}\left(x^{2}\right)^{\frac{D}{2}}}-\frac{\delta_{ab}m_{q}}{4\pi^{2}\left(x^{2}\right)^{\frac{D}{2}-1}}-\frac{\delta_{ab}}{12}\left\langle \bar{q}q\right\rangle +\frac{\mathrm{i}\delta_{ab}x\!\!\!/}{48}m_{q}\left\langle \bar{q}q\right\rangle -\frac{x^{2}\delta_{ab}}{192}\left\langle \bar{q}Gq\right\rangle \nonumber \\
 & +\frac{\mathrm{i}\delta_{ab}x^{2}x\!\!\!/}{1152}m_{q}\left\langle \bar{q}Gq\right\rangle -g_{s}\frac{\mathrm{i}t_{ab}^{N}}{32\pi^{2}\left(x^{2}\right)^{\frac{D}{2}-1}}\left(\sigma^{\alpha\beta}x\!\!\!/+x\!\!\!/\sigma^{\alpha\beta}\right)\nonumber \\
 & -\frac{t_{ab}^{N}\sigma_{\alpha\beta}}{192}\left\langle \bar{q}Gq\right\rangle +\frac{\mathrm{i}t_{ab}^{N}}{768}\left(\sigma_{\alpha\beta}x\!\!\!/+x\!\!\!/\sigma_{\alpha\beta}\right)m_{q}\left\langle \bar{q}Gq\right\rangle\, ,
\end{align}
and that for the heavy quarks
\begin{align}
\mathrm{i}S^{Q}\left(p,m\right) & =\frac{\mathrm{i}\left(p\!\!\!/+m\right)}{\left(p^{2}-m^{2}+\mathrm{i}\epsilon\right)}\delta^{ab}-G_{\mu\nu}^{c}\frac{\mathrm{i}}{4}\left(t^{c}\right)^{ab}\frac{\sigma^{\mu\nu}\left(p\!\!\!/+m\right)+\left(p\!\!\!/+m\right)\sigma^{\mu\nu}}{\left(p^{2}-m+\mathrm{i}\epsilon\right)^{2}}\nonumber \\
 & +\frac{\mathrm{i}}{12}\delta^{ab}\left\langle g^{2}G_{\alpha\beta}^{a}G_{\alpha\beta}^{a}\right\rangle m\frac{p^{2}+mp\!\!\!/}{\left(p^{2}-m^{2}+\mathrm{i}\epsilon\right)^{4}}\, .
\end{align}
The propagator of the light quark is presented in coordinate space,
while the propagator of the heavy quark is presented in momentum
space. These two forms are related by Fourier transformation,
\begin{equation}
S\left(x\right)=\int\frac{\mathrm{d}^{4}p}{\left(2\pi\right)^{4}}e^{-\mathrm{i}p\cdot
x}S\left(p\right)\, .
\end{equation}

In order to suppress the contribution of the higher states and
remove the unknown subtraction terms in Eq. \eqref{eq:dispersion
relation}, we perform the Borel transformation to the correlation
function. The Borel transformation is defined as
\begin{equation}
\hat{B}\left[f\left(q^{2}\right)\right]=\underset{\begin{array}{c}
-q^{2},n\rightarrow\infty\\
-\nicefrac{q^{2}}{n}\equiv M_{B}^{2}
\end{array}}{\lim}\frac{1}{n!}\left(-q^{2}\right)^{n+1}\left(\frac{\mathrm{d}}{\mathrm{d}q^{2}}\right)^{n}f\left(q^{2}\right)
\end{equation}
The correlation function after \chadd{the} Borel transformation can
be expressed as
\begin{equation}
\hat{B}\Pi\left(q^{2}\right)=\frac{1}{\pi}\int
ds\,e^{-\nicefrac{s}{M_{B}^{2}}}\mathrm{Im}\Pi\left(s\right)\, .
\end{equation}
Comparing the correlation function at both \chadd{the} OPE side and
phenomenological side and using quark-hadron duality, one gets
\begin{equation}
f_{X}^{2}e^{-\nicefrac{m_{X}^{2}}{M_{B}^{2}}}=\int_{<}^{s_{0}}dx\,e^{-\nicefrac{s}{M_{B}^{2}}}\rho\left(s\right)\,
, \label{eq:sum rule}
\end{equation}
where $s_{0}$ is the threshold parameter and $M_{B}$ is the Borel
parameter. The expressions of the spectral functions
$\rho\left(s\right)$ are very complicated and lengthy. We only
present the expressions of the spectral functions for $J_3$ and
$J_{3\mu}$ in Appendix~\ref{spectral_fucntion}. Using the sum rule
Eq. (\ref{eq:sum rule}), the mass of the resonance can be extracted
as
\begin{equation}
m^{2}=\frac{\int_{<}^{s_{0}}ds\,s\rho^{OPE}\left(s\right)e^{-\frac{s}{M_{B}^{2}}}}{\int_{<}^{s_{0}}ds\,\rho^{OPE}\left(s\right)e^{-\frac{s}{M_{B}^{2}}}}\,
.
\end{equation}
As a function of $s_{0}$ and $M_{B}$, we need to find suitable
working regions in the parameter space
$\left(s_{0},M_{B}^{2}\right)$ to extract the hadron mass. We will
discuss the details in the next section.

%%%%%%%%%%%%%%%%%%%%%%%%%%%%%%%%%%
\section{Numerical analysis}
\label{sec3}
%%%%%%%%%%%%%%%%%%%%%%%%%%%%%%%%%%

We will use the following parameters
\cite{1985-Reinders-p1-1,2016-Patrignani-p100001-100001,2012-Narison-p259-263,2007-Kuhn-p192-215}
in the numerical analysis
\begin{align}
m_{c}\left(m_{c}\right) & =\bar{m}_{c}=\left(1.275\pm0.025\right)\mathrm{GeV},\nonumber \\
m_{b}\left(m_{b}\right) & =\bar{m}_{b}=\left(4.18\pm0.03\right)\mathrm{GeV},\nonumber \\
\left\langle \bar{q}q\right\rangle  & =-\left(0.24\pm0.01\right)^{3}\mathrm{GeV}^{3},\nonumber \\
\left\langle \bar{q}g_{s}\sigma\cdot Gq\right\rangle  & =-M_{0}^{2}\left\langle \bar{q}q\right\rangle ,\\
M_{0}^{2} & =\left(0.8\pm0.2\right)\mathrm{GeV}^{2},\nonumber \\
\left\langle g_{s}^{2}GG\right\rangle  &
=\left(0.48\pm0.14\right)\mathrm{GeV}^{4}\, .\nonumber
\end{align}
To determine the Borel parameter, we first require that the
contribution of the quark condensate term be less than 30\% of the
contribution of the perturbation term. Such a limitation ensures the
convergence of the OPE series and leads to the lower bound on
$M_{B}$. We then require the following pole contribution to be
larger than 10\%
\begin{equation}
PC\left(s_{0},M_{B}^{2}\right)=\frac{\int_{<}^{s_{0}}ds\,\rho^{OPE}\left(s\right)e^{-\frac{s}{M_{B}^{2}}}}{\int_{<}^{\infty}ds\,\rho^{OPE}\left(s\right)e^{-\frac{s}{M_{B}^{2}}}}\,
,
\end{equation}
which gives the upper bound on $M_{B}$. An optimal value for the
threshold parameter $s_{0}$ is determined by requiring the $M_{B}$
dependence of the hadron mass $m$ as less as possible. These
criteria lead to suitable working regions for $s_{0}$ and $M_{B}$.

\begin{figure}
\begin{centering}
\includegraphics[scale=0.8]{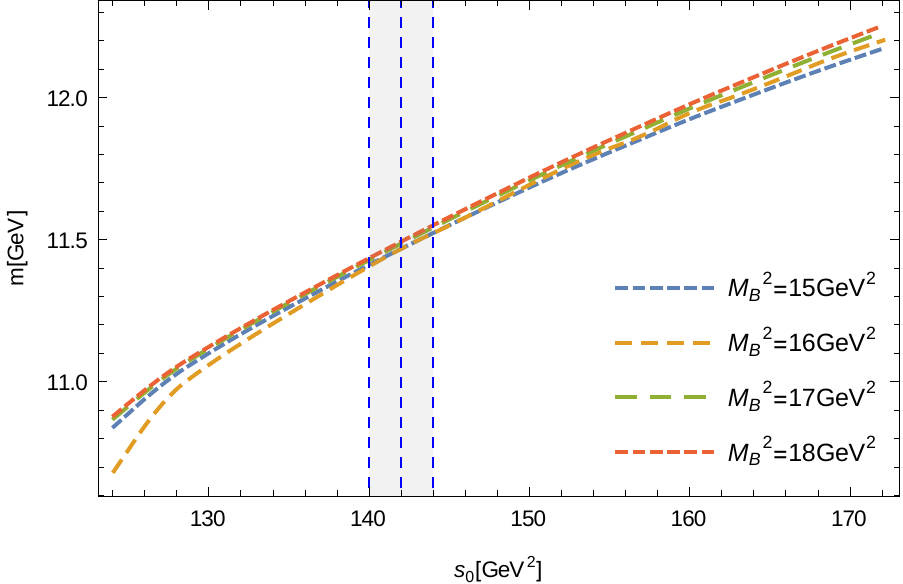}
\includegraphics[scale=0.8]{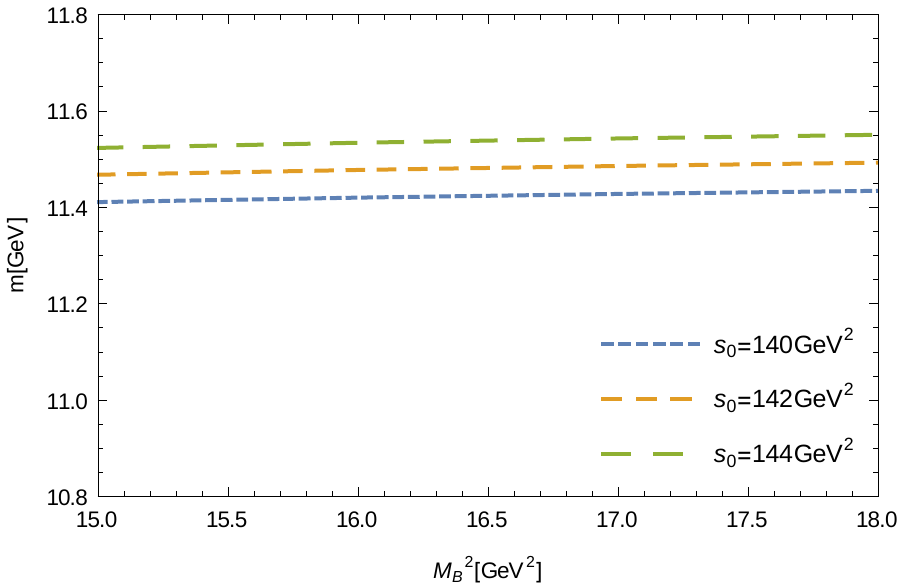}
\par\end{centering}
\caption{\label{fig:The-mass-curve}The dependence of the extracted
mass on the threshold parameter $s_{0}$ and $M_{B}$ for the current
$J_{3}^{bc\bar{b}q}$. }
\end{figure}

We take the current $J_{3}^{bc\bar{c}d}$ as an example to illustrate
the mass sum rule analyses. We show the hadron mass as a function of
the continuum threshold $s_0$ and Borel parameter $M_B^2$ in Fig.
\ref{fig:The-mass-curve}. From the left graph of Fig.
\ref{fig:The-mass-curve}, the $M_{B}$ dependance of the extracted
mass becomes very weak for $140 \mbox{GeV}^{2}\le s_{0}\le 144
\mbox{GeV}^{2}$. In this region, we plot the hadron mass $m$ as a
function of the Borel parameter $M_{B}$ in the right graph of Fig.
\ref{fig:The-mass-curve}. The mass curves are very stable in such a
Borel window. Accordingly, we extract the tetraquark mass as
\begin{equation}
m=11.5\pm0.3~\mathrm{GeV}\, ,
\end{equation}
where the error comes from the errors of the input parameters
$m_{b}$, $m_{c}$, the condensates $\left\langle
\bar{q}q\right\rangle ,\left\langle g_{s}^{2}GG\right\rangle
,\left\langle \bar{q}g_{s}\sigma\cdot Gq\right\rangle $, and the
uncertainties of $M_{B}$ and $s_{0}$.

For the other tetraquark systems, we can perform similar mass sum
rule analyses and extract their masses.
%We show the mass curves for these systems in Appendix~\ref{masscurves}.
The extracted masses for the $cc\bar{b}\bar{q}$, $cc\bar{c}\bar{q}$,
$bc\bar{b}\bar{q}$, $bb\bar{b}\bar{q}$ tetraquarks with
$J^{P}=0^{+}$ and $J^{P}=1^{+}$ are listed in Table
\ref{tab:j0_masses} and Table \ref{tab:j1_masses}, respectively. 
For the $cc\bar c\bar q$ tetraquarks, it is shown that the interpolating 
currents $J_1$ and $J_4$ lead to the same numerical results of 
the hadron mass, although they have totally different diquark-antidiquark 
color structures. In principle, the color anti-triplet diquark is the attractive 
channel while the color sextet diquark is the repulsive one in the 
one-gluon exchange model. However, the interaction between 
the color $6$ and $\bar 6$ diquarks will be attractive when they 
form a color singlet tetraquark. 
In other words, the sum of the repulsion between the two quarks
within the diquark/anti-diquark and the strong attraction between
the 6 and $\bar 6$ diquarks leads to a net attraction, which is roughly
the same as the total attraction in the $3-\bar 3$ channel. Therefore,
we obtain roughly equal masses of the particles associated to $J_1$
and $J_4$ within errors.

\begin{table}
\begin{centering}
\begin{tabular}{|c|c|c|c|c|}
\hline System & Current & $s_{0}\left(\mathrm{GeV}^{2}\right)$ &
$\left[M_{B,min}^{2},M_{B,max}^{2}\right]$ &
$m\left(\mathrm{GeV}\right)$\tabularnewline \hline
\multirow{4}{*}{$bc\bar{b}\bar{q}$} & $j_{1}$ & $133$ &
$\left[15,18\right]$ & $11.4\pm0.6$\tabularnewline \cline{2-5}
 & $j_{2}$ & $136$ & $\left[15,18\right]$ & $11.3\pm0.3$\tabularnewline
\cline{2-5}
 & $j_{3}$ & $142$ & $\left[15,18\right]$ & $11.5\pm0.3$\tabularnewline
\cline{2-5}
 & $j_{4}$ & $141$ & $\left[15,18\right]$ & $11.4\pm0.4$\tabularnewline
\hline \multirow{2}{*}{$cc\bar{b}\bar{q}$} & $j_{1}$ & $70$ &
$\left[8,11\right]$ & $8.0\pm0.3$\tabularnewline \cline{2-5}
 & $j_{4}$ & $73$ & $\left[8,11\right]$ & $8.2\pm0.3$\tabularnewline
\hline \multirow{2}{*}{$cc\bar{c}\bar{q}$} & $j_{1}$ & $29$ &
$\left[5,8\right]$ & $5.1\pm0.2$\tabularnewline \cline{2-5}
 & $j_{4}$ & $30$ & $\left[5,8\right]$ & $5.1\pm0.2$\tabularnewline
\hline \multirow{2}{*}{$bb\bar{b}\bar{q}$} & $j_{1}$ & $189$ &
$\left[16,19\right]$ & $13.5\pm0.4$\tabularnewline \cline{2-5}
 & $j_{4}$ & $198$ & $\left[16,19\right]$ & $13.7\pm0.3$\tabularnewline
\hline
\end{tabular}
\par\end{centering}
\caption{\label{tab:j0_masses}The extracted masses for various
tetraquarks with $J^{P}=0^{+}$. }
\end{table}

\begin{table}

\begin{centering}
\begin{tabular}{|c|c|c|c|c|}
\hline System & Current & $s_{0}\left(\mathrm{GeV}^{2}\right)$ &
$\left[M_{B,min}^{2},M_{B,max}^{2}\right]$ &
$m\left(\mathrm{GeV}\right)$\tabularnewline \hline
\multirow{4}{*}{$bc\bar{b}\bar{q}$} & $j_{1\mu}$ & $142$ &
$\left[15,18\right]$ & $11.5\pm0.5$\tabularnewline \cline{2-5}
 & $j_{2\mu}$ & $136$ & $\left[15,18\right]$ & $11.3\pm0.4$\tabularnewline
\cline{2-5}
 & $j_{3\mu}$ & $144$ & $\left[15,18\right]$ & $11.6\pm0.4$\tabularnewline
\cline{2-5}
 & $j_{4\mu}$ & $140$ & $\left[15,18\right]$ & $11.4\pm0.5$\tabularnewline
\hline \multirow{2}{*}{$cc\bar{b}\bar{q}$} & $j_{1\mu}$ & $71$ &
$\left[8,11\right]$ & $8.1\pm0.3$\tabularnewline \cline{2-5}
 & $j_{4\mu}$ & $73$ & $\left[8,11\right]$ & $8.2\pm0.3$\tabularnewline
\hline \multirow{2}{*}{$cc\bar{c}\bar{q}$} & $j_{1\mu}$ & $30$ &
$\left[5,8\right]$ & $5.1\pm0.2$\tabularnewline \cline{2-5}
 & $j_{4\mu}$ & $30$ & $\left[5,8\right]$ & $5.1\pm0.2$\tabularnewline
\hline \multirow{2}{*}{$bb\bar{b}\bar{q}$} & $j_{1\mu}$ & $189$ &
$\left[16,19\right]$ & $13.5\pm0.4$\tabularnewline \cline{2-5}
 & $j_{4\mu}$ & $189$ & $\left[16,19\right]$ & $13.5\pm0.4$\tabularnewline
\hline
\end{tabular}
\par\end{centering}
\caption{\label{tab:j1_masses}The extracted masses for various
tetraquarks with $J^{P}=1^{+}$.}

\end{table}

In the following, we discuss the decay patterns of the possible
$cc\bar{b}\bar{q}$, $cc\bar{c}\bar{q}$, $bc\bar{b}\bar{q}$,
$bb\bar{b}\bar{q}$ tetraquark states using the mass spectra obtained
in the previous section. We will only consider the two-body hadronic
decays. In fact, these tetraquark states may decay into meson pairs
easily if the kinematics allows.

Considering the conservations of the spin-parity, we list the
possible decay modes for the $cc\bar{c}\bar{q}$, $cc\bar{b}\bar{q}$,
$bc\bar{b}\bar{q}$, $bb\bar{b}\bar{q}$ tetraquarks in
Table~\ref{tab:Decay-modes}. From Tables
\ref{tab:j0_masses}-\ref{tab:j1_masses}, the masses of the
$cc\bar{c}\bar{q}$, $cc\bar{b}\bar{q}$, $bc\bar{b}\bar{q}$
tetraquarks are higher than the corresponding meson-meson
thresholds. They will decay easily through rearrangement or the
so-called fall-apart mechanism. Since the decay momentum is not very
large, these tetraquarks are not expected to be very broad
resonances. Thus it may be difficult and still possible to observe
them experimentally. However, the situation is different for the
$bb\bar{b}\bar{q}$ tetraquark systems. Note that the extracted
masses for both the scalar and axial-vector triply bottomed
tetraquarks are lower than the open bottom thresholds
$T_{\eta_{b}B}=14.68\mathrm{GeV}$ and
$T_{\eta_{b}B^{*}}=14.72\mathrm{GeV}$. These tetraquark states
cannot decay into a bottomonia plus a $B^{(*)}$ meson, which implies
that the $bb\bar{b}\bar{q}$ states should be stable.

\begin{center}
\begin{table}
\begin{centering}
\begin{tabular}{|c|c|c|}
\hline $J^{P}$ & Flavor content & S-wave decay\tabularnewline \hline
\multirow{4}{*}{$0^{+}$} & $cc\bar{c}\bar{q}$ & $\left(J/\psi
D^{*}\right),\left(\eta_{c}D\right)$\tabularnewline \cline{2-3}
 & $cc\bar{b}\bar{q}$ & $\left(B_{c}D\right)$\tabularnewline
\cline{2-3}
 & $bc\bar{b}\bar{q}$ & $\left(BB_{c}\right),\left(\varUpsilon D^{*}\right),\left(\eta_{b}D\right)$\tabularnewline
\cline{2-3}
 & $bb\bar{b}\bar{q}$ & -\tabularnewline
\hline \multirow{4}{*}{$1^{+}$} & $cc\bar{c}\bar{q}$ & $\left(J/\psi
D^{*}\right),\left(\eta_{c}D^{*}\right),\left(J/\psi
D\right)$\tabularnewline \cline{2-3}
 & $cc\bar{b}\bar{q}$ & $\left(B_{c}D^{*}\right)$\tabularnewline
\cline{2-3}
 & $bc\bar{b}\bar{q}$ & $\left(\varUpsilon D^{*}\right),\left(\eta_{b}D^{*}\right),\left(\varUpsilon D\right)$\tabularnewline
\cline{2-3}
 & $bb\bar{b}\bar{q}$ & -\tabularnewline
\hline
\end{tabular}
\par\end{centering}
\caption{\label{tab:Decay-modes}Decay modes of the tetraquark states
with different quantum numbers and flavor contents.}
\end{table}
\par\end{center}

%%%%%%%%%%%%%%%%%%%%%%%%%%%%%%%%%%
\section{Discussions and Conclusions}
\label{sec5}
%%%%%%%%%%%%%%%%%%%%%%%%%%%%%%%%%%

In the framework of QCD sum rules, we have investigated the mass
spectra of the triply heavy tetraquark states with $J^{P}=0^{+}$ and
$J^{P}=1^{+}$. We first construct the interpolating currents of the
$cc\bar{c}\bar{q}$, $cc\bar{b}\bar{q}$, $bc\bar{b}\bar{q}$,
$bb\bar{b}\bar{q}$ systems with quantum numbers $J^{P}=0^{+},
1^{+}$. Using these interpolating currents, we have calculated the
two-point correlation functions and spectral densities. Then we
perform numerical analyses for the mass sum rules and extract the
mass spectra of various triply heavy tetraquark states. Accordingly,
we have also discussed the possible decay patterns of these
tetraquark states.

As shown in Tables \ref{tab:j0_masses}-\ref{tab:j1_masses}, the
masses of the $cc\bar{c}\bar{q}$, $cc\bar{b}\bar{q}$,
$bc\bar{b}\bar{q}$ tetraquarks seem higher than some two-meson
thresholds. They may decay easily into the two-body final states
through the fall-apart mechanism. However, these results are much
lower than those obtained in the framework of the color-magnetic
interaction~\cite{2017-Chen-p5-5}. Due to the limited phase space,
these states  might not be very broad. In contrast, the
$bb\bar{b}\bar{q}$ tetraquark states lie below the bottomonia plus
$B^{(*)}$ thresholds and their OZI-allowed strong decays are
kinematically forbidden. It is very interesting to note that some
$bb\bar{q}\bar{q}$ states also lie below two bottom meson threshold
and are probably stable~\cite{2013-Du-p14003-14003,2014-Chen-p54037-54037,
2017-Karliner-p-,2007-Cui-p7-13,2017-Luo-p-}.

However, the heavy quark pair annihilation and light quark pair
creation processes $bb\bar b\bar q\to (b\bar q) + (q\bar q)$ are
still possible. Such characteristic $B$ meson plus a light meson
modes will contribute significantly to the $bb\bar{b}\bar{q}$ decay
width. These $bb\bar{b}\bar{q}$ tetraquark states may also decay via
electromagnetic and weak interactions. They can decay into $\bar
B^{(\ast)}\gamma$ through the $b\gamma_\mu\bar b\to\gamma$ photon
production process. The weak decay $bb\bar b\bar q\to J/\psi\Upsilon
K$ is also allowed, although the phase space is limited. If such
$bb\bar{b}\bar{q}$ states do exist, they may be detected in the
final states with $B$ meson and other light mesons or photon. With
the running of LHC at 13 TeV and the forthcoming BelleII, searching
for such triply heavy tetraquark states will probably become
feasible in the near future.

\section*{ACKNOWLEDGMENTS}

This project is supported by the National Natural Science Foundation
of China under Grants 11575008, 11621131001 and 973 program. 
Wei Chen is supported in part by the Chinese National Youth Thousand 
Talents Program.

\clearpage
\appendix

\section{The spectral density functions} \label{spectral_fucntion}

For the interpolating current $J_{3}$, the spectral functions are

{\scriptsize{}
\begin{equation}
\rho^{OPE}\left(s\right)=\rho^{pert}\left(s\right)+\rho^{\left\langle
\bar{q}q\right\rangle }\left(s\right)+\rho^{\left\langle
GG\right\rangle }\left(s\right)+\rho^{\left\langle
\bar{q}Gq\right\rangle }\left(s\right)
\end{equation}
\begin{align}
 & \rho^{pert}\left(s\right)=\frac{1}{\pi}\mathrm{Im}\Pi^{pert}\left(s\right)\nonumber \\
 & =\int_{0}^{1}d\alpha\int_{0}^{1-\alpha}d\beta\int_{0}^{1-\alpha-\beta}d\gamma\theta\left(s-\frac{\text{m}_{1}^{2}}{\alpha}-\frac{\text{m}_{2}^{2}}{\beta}-\frac{\text{m}_{2}^{2}}{\gamma}\right)\frac{-\alpha\beta}{256\pi^{6}}\left(\frac{\text{m}_{1}^{2}}{\alpha}+\frac{\text{m}_{2}^{2}}{\beta}+\frac{\text{m}_{2}^{2}}{\gamma}-s\right)^{2}\frac{1}{\left(\alpha+\beta-1\right)}\times\nonumber \\
 & \biggl\{3\gamma(\alpha+\beta-1)^{2}(\alpha+\beta+\gamma-1)\left(\frac{\text{m1}^{2}}{\alpha}+\frac{\text{m2}^{2}}{\beta}+\frac{\text{m3}^{2}}{\gamma}-s\right)^{2}\nonumber \\
 & -2s\left(\frac{\text{m1}^{2}}{\alpha}+\frac{\text{m2}^{2}}{\beta}+\frac{\text{m3}^{2}}{\gamma}-s\right)\biggl[5\alpha^{4}+\alpha^{3}(13\beta+12\gamma-13)+\alpha^{2}\left(11\beta^{2}+\beta(30\gamma-22)+6\gamma^{2}-23\gamma+11\right)\nonumber \\
 & +\alpha\left(3\beta^{3}+3\beta^{2}(8\gamma-3)+\beta\left(12\gamma^{2}-38\gamma+9\right)-6\gamma^{2}+14\gamma-3\right)+\gamma\left(6\beta^{3}+3\beta^{2}(2\gamma-5)-6\beta(\gamma-2)+2\gamma-3\right)\biggr]\nonumber \\
 & +6s^{2}\left(\alpha^{2}+\alpha(\beta+\gamma-1)+\beta\gamma\right)\left(2\alpha^{2}+\alpha(3\beta+\gamma-3)+\beta^{2}+\beta(\gamma-2)+1\right)\biggl\}\nonumber \\
 & +\int d\alpha\int d\beta\int d\gamma\frac{1}{128\pi^{6}}\text{m1}\text{m2}\left(\frac{\text{m1}^{2}}{\alpha}+\frac{\text{m2}^{2}}{\beta}+\frac{\text{m3}^{2}}{\gamma}-s\right)\times\nonumber \\
 & \biggl[2(\alpha+\beta+\gamma-1)\gamma\left(\frac{\text{m1}^{2}}{\alpha}+\frac{\text{m2}^{2}}{\beta}+\frac{\text{m3}^{2}}{\gamma}-s\right)\nonumber \\
 & -\frac{3s\left(\alpha^{2}+\alpha(\beta+\gamma-1)+\beta\gamma\right)\left(2\alpha^{2}+\alpha(3\beta+\gamma-3)+\beta^{2}+\beta(\gamma-2)+1\right)}{(\alpha+\beta-1)^{2}}\biggr]^{2}
\end{align}
\begin{align}
 & \rho^{\left\langle \bar{q}q\right\rangle }\left(s\right)=\frac{1}{\pi}\mathrm{Im}\Pi^{\left\langle \bar{q}q\right\rangle }\left(q^{2}\right)\nonumber \\
 & =\left\langle \bar{q}q\right\rangle \int_{0}^{1}{\rm d}\alpha\int_{0}^{1-\alpha}d\beta\theta\left(s-\frac{m_{1}^{2}}{\alpha}-\frac{m_{2}^{2}}{\beta}-\frac{m_{3}^{2}}{1-\alpha-\beta}\right)\biggl[\frac{m_{1}m_{2}m_{3}}{16\pi^{4}}\left(\frac{m_{1}^{2}}{\alpha}+\frac{m_{2}^{2}}{\beta}+\frac{m_{3}^{2}}{1-\alpha-\beta}-s\right)\nonumber \\
 & -\frac{m_{3}\left[\left(\alpha+\beta-1\right)\left(\alpha m_{2}^{2}+\beta m_{1}^{2}-\alpha\beta s\right)-\alpha\beta m_{3}^{2}\right]\left[\left(\alpha+\beta-1\right)\left(\alpha m_{2}^{2}+\beta m_{1}^{2}-2\alpha\beta s\right)-\alpha\beta m_{3}^{2}\right]}{16\pi^{4}\alpha\beta\left(1-\alpha-\beta\right)^{2}}\biggr]
\end{align}
}{\tiny{}
\begin{align}
 & \rho^{\left\langle GG\right\rangle }\left(s\right)=\nonumber \\
 & \int_{0}^{1}d\alpha\int_{0}^{1-\alpha}d\beta\int_{0}^{1-\alpha-\beta}d\gamma\theta\left(s-\frac{\text{m}_{1}^{2}}{\alpha}-\frac{\text{m}_{2}^{2}}{\beta}-\frac{\text{m}_{2}^{2}}{\gamma}\right)\frac{-\left\langle g_{s}^{2}GG\right\rangle }{3072\pi^{6}}\biggl\{-\frac{4\gamma\text{m1}^{3}\text{m2}(\alpha+\beta+\gamma-1)}{\alpha^{3}}\nonumber \\
 & -\frac{4\gamma\text{m1}\text{m2}^{3}(\alpha+\beta+\gamma-1)}{\beta^{3}}-\frac{4\text{m1}\text{m2}\text{m3}^{2}(\alpha+\beta+\gamma-1)}{\gamma^{2}}\nonumber \\
 & +\frac{1}{\alpha^{3}(\alpha+\beta-1)^{2}}2\text{m1}^{2}\biggl[6\gamma(\alpha+\beta-1)^{2}(\alpha+\beta+\gamma-1)\left(\beta\text{m1}^{2}+\alpha\text{m2}^{2}\right)+6\alpha\beta\text{m3}^{2}(\alpha+\beta-1)^{2}(\alpha+\beta+\gamma-1)\nonumber \\
 & -\alpha\beta s\biggl(5\alpha^{4}+\alpha^{3}(13\beta+18\gamma-13)+\alpha^{2}\left(11\beta^{2}+\beta(48\gamma-22)+12\gamma^{2}-41\gamma+11\right)\nonumber \\
 & +\alpha\left(3\beta^{3}+\beta^{2}(42\gamma-9)+\beta\left(24\gamma^{2}-74\gamma+9\right)-18\gamma^{2}+32\gamma-3\right)+\gamma\left(12\beta^{3}+3\beta^{2}(4\gamma-11)-6\beta(3\gamma-5)+8\gamma-9\right)\biggr)\biggr]\nonumber \\
 & -\frac{1}{\alpha^{3}\beta^{3}\gamma(\alpha+\beta-1)^{2}(\alpha+\beta+\gamma-1)}3\text{m2}\left(2\alpha^{3}\gamma+\alpha^{2}\left(\beta^{2}+2(\gamma-1)\gamma\right)-2\alpha\beta(\gamma-1)\gamma+2\beta^{2}\gamma(\beta+\gamma-1)\right)\times\nonumber \\
 & \biggl[2\gamma(\alpha+\beta-1)^{2}(\alpha+\beta+\gamma-1)\left(\beta\text{m1}^{2}+\alpha\text{m2}^{2}\right)+2\alpha\beta\text{m3}^{2}(\alpha+\beta-1)^{2}(\alpha+\beta+\gamma-1)-\alpha\beta s\nonumber \\
 & \biggl(2\alpha^{4}+5\alpha^{3}(\beta+\gamma-1)+\alpha^{2}\left(4\beta^{2}+\beta(13\gamma-8)+3\gamma^{2}-10\gamma+4\right)\nonumber \\
 & +\alpha\left(\beta^{3}+\beta^{2}(11\gamma-3)+3\beta\left(2\gamma^{2}-6\gamma+1\right)-4\gamma^{2}+7\gamma-1\right)+\gamma\left(3\beta^{3}+\beta^{2}(3\gamma-8)+\beta(7-4\gamma)+2(\gamma-1)\right)\biggr)\biggr]\nonumber \\
 & +\biggl[\frac{s^{2}\left(2\alpha^{4}+\alpha^{3}(5\beta+3\gamma-5)+\alpha^{2}\left(4\beta^{2}+\beta(7\gamma-8)+(\gamma-2)^{2}\right)+\alpha\left(\beta^{3}+\beta^{2}(5\gamma-3)+\beta\left(2\gamma^{2}-6\gamma+3\right)+\gamma-1\right)+\beta\gamma\left(\beta^{2}+\beta(\gamma-2)+1\right)\right)}{(\alpha+\beta-1)^{2}}\nonumber \\
 & +\frac{s}{\alpha\beta\gamma(\alpha+\beta-1)^{2}}\left(-\gamma\left(\beta\left(\text{m1}^{2}-\alpha s\right)+\alpha\text{m2}^{2}\right)-\alpha\beta\text{m3}^{2}\right)\times\nonumber \\
 & \biggl(5\alpha^{4}+\alpha^{3}(13\beta+12\gamma-13)+\alpha^{2}\left(11\beta^{2}+\beta(30\gamma-22)+6\gamma^{2}-23\gamma+11\right)\nonumber \\
 & +\alpha\left(3\beta^{3}+3\beta^{2}(8\gamma-3)+\beta\left(12\gamma^{2}-38\gamma+9\right)-6\gamma^{2}+14\gamma-3\right)+\gamma\left(6\beta^{3}+3\beta^{2}(2\gamma-5)-6\beta(\gamma-2)+2\gamma-3\right)\biggr)\nonumber \\
 & +\frac{3(\alpha+\beta+\gamma-1)\left(\gamma\left(\beta\left(\text{m1}^{2}-\alpha s\right)+\alpha\text{m2}^{2}\right)+\alpha\beta\text{m3}^{2}\right)^{2}}{\alpha^{2}\beta^{2}\gamma}\biggr]\nonumber \\
 & +\frac{2}{(\alpha+\beta-1)^{2}}\left(\frac{\text{m2}^{2}}{\beta^{3}}+\frac{\text{m3}^{2}}{\gamma^{3}}\right)\times\biggl[6\gamma(\alpha+\beta-1)^{2}(\alpha+\beta+\gamma-1)\left(\beta\text{m1}^{2}+\alpha\text{m2}^{2}\right)\nonumber \\
 & +6\alpha\beta\text{m3}^{2}(\alpha+\beta-1)^{2}(\alpha+\beta+\gamma-1)-\alpha\beta s\biggl(5\alpha^{4}+\alpha^{3}(13\beta+18\gamma-13)+\alpha^{2}\left(11\beta^{2}+\beta(48\gamma-22)+12\gamma^{2}-41\gamma+11\right)\nonumber \\
 & +\alpha\left(3\beta^{3}+\beta^{2}(42\gamma-9)+\beta\left(24\gamma^{2}-74\gamma+9\right)-18\gamma^{2}+32\gamma-3\right)+\gamma\left(12\beta^{3}+3\beta^{2}(4\gamma-11)-6\beta(3\gamma-5)+8\gamma-9\right)\biggr)\biggr]\biggr\}\nonumber \\
 & +\int_{0}^{1}d\alpha\int_{0}^{1-\alpha}d\beta\int_{0}^{1-\alpha-\beta}d\gamma\delta\left(s-\frac{m_{1}^{2}}{\alpha}-\frac{m_{2}^{2}}{\beta}-\frac{m_{3}^{2}}{1-\alpha-\beta}\right)\frac{1}{1536\pi^{6}\alpha^{3}\beta^{3}\gamma^{3}(\alpha+\beta-1)^{2}}\biggl\{\left(\gamma^{3}\left(\beta^{3}\text{m1}^{2}+\alpha^{3}\text{m2}^{2}\right)+\alpha^{3}\beta^{3}\text{m3}^{2}\right)\times\nonumber \\
 & \left(\text{m1}\text{m2}+\alpha\beta s\right)\biggl[2\alpha^{4}+\alpha^{3}(5\beta+3\gamma-5)+\alpha^{2}\left(4\beta^{2}+\beta(7\gamma-8)+(\gamma-2)^{2}\right)\nonumber \\
 & +\alpha\left(\beta^{3}+\beta^{2}(5\gamma-3)+\beta\left(2\gamma^{2}-6\gamma+3\right)+\gamma-1\right)+\beta\gamma\left(\beta^{2}+\beta(\gamma-2)+1\right)\biggr]\biggr\}
\end{align}
}{\tiny \par}

{\scriptsize{}
\begin{align}
 & \rho^{\left\langle \bar{q}Gq\right\rangle }\left(s\right)=\frac{1}{\pi}\mathrm{Im}\Pi^{\left\langle \bar{q}Gq\right\rangle }\left(q^{2}\right)=\nonumber \\
 & \left\langle \bar{q}Gq\right\rangle \int_{0}^{1}{\rm d}\alpha\int_{0}^{1-\alpha}d\beta\frac{-\text{m3}s(\text{m1}\text{m2}+\alpha\beta s)}{64\pi^{4}}\delta\left(s-\frac{m_{1}^{2}}{\alpha}-\frac{m_{2}^{2}}{\beta}-\frac{m_{3}^{2}}{1-\alpha-\beta}\right)\nonumber \\
 & +\int_{0}^{1}{\rm d}\alpha\int_{0}^{1-\alpha}d\beta\theta\left(s-\frac{m_{1}^{2}}{\alpha}-\frac{m_{2}^{2}}{\beta}-\frac{m_{3}^{2}}{1-\alpha-\beta}\right)\frac{-\text{m3}}{64\pi^{4}(\alpha+\beta-1)^{2}}\biggl[-2\beta\text{m1}^{2}\left(3\alpha^{2}+\alpha(6\beta-7)+3\beta^{2}-7\beta+4\right)\nonumber \\
 & +\text{m1}\text{m2}(2\alpha+2\beta-3)(\alpha+\beta-1)+\alpha\biggl(\beta\left(2\text{m3}^{2}(3\alpha+3\beta-4)+3s(4\alpha+4\beta-5)(\alpha+\beta-1)\right)\nonumber \\
 & -2\text{m2}^{2}\left(3\alpha^{2}+\alpha(6\beta-7)+3\beta^{2}-7\beta+4\right)\biggr)\biggr]
\end{align}
}

For the interpolating current $J_{3\mu}$, the spectral functions are

\begin{equation}
\rho^{OPE}\left(s\right)=\rho^{pert}\left(s\right)+\rho^{\left\langle
\bar{q}q\right\rangle }\left(s\right)+\rho^{\left\langle
GG\right\rangle }\left(s\right)+\rho^{\left\langle
\bar{q}Gq\right\rangle }\left(s\right)
\end{equation}
{\scriptsize{}
\begin{align}
 & \rho^{pert}\left(s\right)=\frac{1}{\pi}\mathrm{Im}\Pi^{pert}\left(s\right)\nonumber \\
 & =\int_{0}^{1}d\alpha\int_{0}^{1-\alpha}d\beta\int_{0}^{1-\alpha-\beta}d\gamma\theta\left(s-\frac{\text{m}_{1}^{2}}{\alpha}-\frac{\text{m}_{2}^{2}}{\beta}-\frac{\text{m}_{2}^{2}}{\gamma}\right)\frac{-\alpha\beta}{256\pi^{6}}\left(\frac{\text{m}_{1}^{2}}{\alpha}+\frac{\text{m}_{2}^{2}}{\beta}+\frac{\text{m}_{2}^{2}}{\gamma}-s\right)^{2}\frac{1}{\left(\alpha+\beta-1\right)}\times\nonumber \\
 & \biggl\{3\gamma(\alpha+\beta-1)^{2}(\alpha+\beta+\gamma-1)\left(\frac{\text{m1}^{2}}{\alpha}+\frac{\text{m2}^{2}}{\beta}+\frac{\text{m3}^{2}}{\gamma}-s\right)^{2}\nonumber \\
 & -2s\left(\frac{\text{m1}^{2}}{\alpha}+\frac{\text{m2}^{2}}{\beta}+\frac{\text{m3}^{2}}{\gamma}-s\right)\biggl[5\alpha^{4}+\alpha^{3}(13\beta+12\gamma-13)+\alpha^{2}\left(11\beta^{2}+\beta(30\gamma-22)+6\gamma^{2}-23\gamma+11\right)\nonumber \\
 & +\alpha\left(3\beta^{3}+3\beta^{2}(8\gamma-3)+\beta\left(12\gamma^{2}-38\gamma+9\right)-6\gamma^{2}+14\gamma-3\right)+\gamma\left(6\beta^{3}+3\beta^{2}(2\gamma-5)-6\beta(\gamma-2)+2\gamma-3\right)\biggr]\nonumber \\
 & +6s^{2}\left(\alpha^{2}+\alpha(\beta+\gamma-1)+\beta\gamma\right)\left(2\alpha^{2}+\alpha(3\beta+\gamma-3)+\beta^{2}+\beta(\gamma-2)+1\right)\biggl\}\nonumber \\
 & +\int d\alpha\int d\beta\int d\gamma\frac{1}{128\pi^{6}}\text{m1}\text{m2}\left(\frac{\text{m1}^{2}}{\alpha}+\frac{\text{m2}^{2}}{\beta}+\frac{\text{m3}^{2}}{\gamma}-s\right)\times\nonumber \\
 & \biggl[2(\alpha+\beta+\gamma-1)\gamma\left(\frac{\text{m1}^{2}}{\alpha}+\frac{\text{m2}^{2}}{\beta}+\frac{\text{m3}^{2}}{\gamma}-s\right)\nonumber \\
 & -\frac{3s\left(\alpha^{2}+\alpha(\beta+\gamma-1)+\beta\gamma\right)\left(2\alpha^{2}+\alpha(3\beta+\gamma-3)+\beta^{2}+\beta(\gamma-2)+1\right)}{(\alpha+\beta-1)^{2}}\biggr]^{2}
\end{align}
\begin{align}
 & \rho^{\left\langle \bar{q}q\right\rangle }\left(s\right)=\frac{1}{\pi}\mathrm{Im}\Pi^{\left\langle \bar{q}q\right\rangle }\left(q^{2}\right)\nonumber \\
 & =\left\langle \bar{q}q\right\rangle \int_{0}^{1}{\rm d}\alpha\int_{0}^{1-\alpha}d\beta\theta\left(s-\frac{m_{1}^{2}}{\alpha}-\frac{m_{2}^{2}}{\beta}-\frac{m_{3}^{2}}{1-\alpha-\beta}\right)\biggl[\frac{m_{1}m_{2}m_{3}}{16\pi^{4}}\left(\frac{m_{1}^{2}}{\alpha}+\frac{m_{2}^{2}}{\beta}+\frac{m_{3}^{2}}{1-\alpha-\beta}-s\right)\nonumber \\
 & -\frac{m_{3}\left[\left(\alpha+\beta-1\right)\left(\alpha m_{2}^{2}+\beta m_{1}^{2}-\alpha\beta s\right)-\alpha\beta m_{3}^{2}\right]\left[\left(\alpha+\beta-1\right)\left(\alpha m_{2}^{2}+\beta m_{1}^{2}-2\alpha\beta s\right)-\alpha\beta m_{3}^{2}\right]}{16\pi^{4}\alpha\beta\left(1-\alpha-\beta\right)^{2}}\biggr]
\end{align}
}{\tiny{}
\begin{align}
 & \rho^{\left\langle GG\right\rangle }\left(s\right)=\nonumber \\
 & \int_{0}^{1}d\alpha\int_{0}^{1-\alpha}d\beta\int_{0}^{1-\alpha-\beta}d\gamma\theta\left(s-\frac{\text{m}_{1}^{2}}{\alpha}-\frac{\text{m}_{2}^{2}}{\beta}-\frac{\text{m}_{2}^{2}}{\gamma}\right)\frac{-\left\langle g_{s}^{2}GG\right\rangle }{3072\pi^{6}}\biggl\{-\frac{4\gamma\text{m1}^{3}\text{m2}(\alpha+\beta+\gamma-1)}{\alpha^{3}}-\frac{4\gamma\text{m1}\text{m2}^{3}(\alpha+\beta+\gamma-1)}{\beta^{3}}-\frac{4\text{m1}\text{m2}\text{m3}^{2}(\alpha+\beta+\gamma-1)}{\gamma^{2}}\nonumber \\
 & +\frac{1}{\alpha^{3}(\alpha+\beta-1)^{2}}2\text{m1}^{2}\biggl[6\gamma(\alpha+\beta-1)^{2}(\alpha+\beta+\gamma-1)\left(\beta\text{m1}^{2}+\alpha\text{m2}^{2}\right)+6\alpha\beta\text{m3}^{2}(\alpha+\beta-1)^{2}(\alpha+\beta+\gamma-1)\nonumber \\
 & -\alpha\beta s\biggl(5\alpha^{4}+\alpha^{3}(13\beta+18\gamma-13)+\alpha^{2}\left(11\beta^{2}+\beta(48\gamma-22)+12\gamma^{2}-41\gamma+11\right)\nonumber \\
 & +\alpha\left(3\beta^{3}+\beta^{2}(42\gamma-9)+\beta\left(24\gamma^{2}-74\gamma+9\right)-18\gamma^{2}+32\gamma-3\right)+\gamma\left(12\beta^{3}+3\beta^{2}(4\gamma-11)-6\beta(3\gamma-5)+8\gamma-9\right)\biggr)\biggr]\nonumber \\
 & -\frac{1}{\alpha^{3}\beta^{3}\gamma(\alpha+\beta-1)^{2}(\alpha+\beta+\gamma-1)}3\text{m2}\left(2\alpha^{3}\gamma+\alpha^{2}\left(\beta^{2}+2(\gamma-1)\gamma\right)-2\alpha\beta(\gamma-1)\gamma+2\beta^{2}\gamma(\beta+\gamma-1)\right)\times\nonumber \\
 & \biggl[2\gamma(\alpha+\beta-1)^{2}(\alpha+\beta+\gamma-1)\left(\beta\text{m1}^{2}+\alpha\text{m2}^{2}\right)+2\alpha\beta\text{m3}^{2}(\alpha+\beta-1)^{2}(\alpha+\beta+\gamma-1)-\alpha\beta s\nonumber \\
 & \biggl(2\alpha^{4}+5\alpha^{3}(\beta+\gamma-1)+\alpha^{2}\left(4\beta^{2}+\beta(13\gamma-8)+3\gamma^{2}-10\gamma+4\right)\nonumber \\
 & +\alpha\left(\beta^{3}+\beta^{2}(11\gamma-3)+3\beta\left(2\gamma^{2}-6\gamma+1\right)-4\gamma^{2}+7\gamma-1\right)+\gamma\left(3\beta^{3}+\beta^{2}(3\gamma-8)+\beta(7-4\gamma)+2(\gamma-1)\right)\biggr)\biggr]\nonumber \\
 & +\biggl[\frac{s^{2}\left(2\alpha^{4}+\alpha^{3}(5\beta+3\gamma-5)+\alpha^{2}\left(4\beta^{2}+\beta(7\gamma-8)+(\gamma-2)^{2}\right)+\alpha\left(\beta^{3}+\beta^{2}(5\gamma-3)+\beta\left(2\gamma^{2}-6\gamma+3\right)+\gamma-1\right)+\beta\gamma\left(\beta^{2}+\beta(\gamma-2)+1\right)\right)}{(\alpha+\beta-1)^{2}}\nonumber \\
 & +\frac{s}{\alpha\beta\gamma(\alpha+\beta-1)^{2}}\left(-\gamma\left(\beta\left(\text{m1}^{2}-\alpha s\right)+\alpha\text{m2}^{2}\right)-\alpha\beta\text{m3}^{2}\right)\times\nonumber \\
 & \biggl(5\alpha^{4}+\alpha^{3}(13\beta+12\gamma-13)+\alpha^{2}\left(11\beta^{2}+\beta(30\gamma-22)+6\gamma^{2}-23\gamma+11\right)\nonumber \\
 & +\alpha\left(3\beta^{3}+3\beta^{2}(8\gamma-3)+\beta\left(12\gamma^{2}-38\gamma+9\right)-6\gamma^{2}+14\gamma-3\right)+\gamma\left(6\beta^{3}+3\beta^{2}(2\gamma-5)-6\beta(\gamma-2)+2\gamma-3\right)\biggr)\nonumber \\
 & +\frac{3(\alpha+\beta+\gamma-1)\left(\gamma\left(\beta\left(\text{m1}^{2}-\alpha s\right)+\alpha\text{m2}^{2}\right)+\alpha\beta\text{m3}^{2}\right)^{2}}{\alpha^{2}\beta^{2}\gamma}\biggr]\nonumber \\
 & +\frac{2}{(\alpha+\beta-1)^{2}}\left(\frac{\text{m2}^{2}}{\beta^{3}}+\frac{\text{m3}^{2}}{\gamma^{3}}\right)\times\biggl[6\gamma(\alpha+\beta-1)^{2}(\alpha+\beta+\gamma-1)\left(\beta\text{m1}^{2}+\alpha\text{m2}^{2}\right)\nonumber \\
 & +6\alpha\beta\text{m3}^{2}(\alpha+\beta-1)^{2}(\alpha+\beta+\gamma-1)-\alpha\beta s\biggl(5\alpha^{4}+\alpha^{3}(13\beta+18\gamma-13)+\alpha^{2}\left(11\beta^{2}+\beta(48\gamma-22)+12\gamma^{2}-41\gamma+11\right)\nonumber \\
 & +\alpha\left(3\beta^{3}+\beta^{2}(42\gamma-9)+\beta\left(24\gamma^{2}-74\gamma+9\right)-18\gamma^{2}+32\gamma-3\right)+\gamma\left(12\beta^{3}+3\beta^{2}(4\gamma-11)-6\beta(3\gamma-5)+8\gamma-9\right)\biggr)\biggr]\biggr\}\nonumber \\
 & +\int_{0}^{1}d\alpha\int_{0}^{1-\alpha}d\beta\int_{0}^{1-\alpha-\beta}d\gamma\delta\left(s-\frac{m_{1}^{2}}{\alpha}-\frac{m_{2}^{2}}{\beta}-\frac{m_{3}^{2}}{1-\alpha-\beta}\right)\frac{1}{1536\pi^{6}\alpha^{3}\beta^{3}\gamma^{3}(\alpha+\beta-1)^{2}}\biggl\{\left(\gamma^{3}\left(\beta^{3}\text{m1}^{2}+\alpha^{3}\text{m2}^{2}\right)+\alpha^{3}\beta^{3}\text{m3}^{2}\right)\times\nonumber \\
 & \left(\text{m1}\text{m2}+\alpha\beta s\right)\biggl[2\alpha^{4}+\alpha^{3}(5\beta+3\gamma-5)+\alpha^{2}\left(4\beta^{2}+\beta(7\gamma-8)+(\gamma-2)^{2}\right)\nonumber \\
 & +\alpha\left(\beta^{3}+\beta^{2}(5\gamma-3)+\beta\left(2\gamma^{2}-6\gamma+3\right)+\gamma-1\right)+\beta\gamma\left(\beta^{2}+\beta(\gamma-2)+1\right)\biggr]\biggr\}
\end{align}
}{\tiny \par}

{\scriptsize{}
\begin{align}
 & \rho^{\left\langle \bar{q}Gq\right\rangle }\left(s\right)=\frac{1}{\pi}\mathrm{Im}\Pi^{\left\langle \bar{q}Gq\right\rangle }\left(q^{2}\right)=\nonumber \\
 & \left\langle \bar{q}Gq\right\rangle \int_{0}^{1}{\rm d}\alpha\int_{0}^{1-\alpha}d\beta\frac{-\text{m3}s(\text{m1}\text{m2}+\alpha\beta s)}{64\pi^{4}}\delta\left(s-\frac{m_{1}^{2}}{\alpha}-\frac{m_{2}^{2}}{\beta}-\frac{m_{3}^{2}}{1-\alpha-\beta}\right)\nonumber \\
 & +\int_{0}^{1}{\rm d}\alpha\int_{0}^{1-\alpha}d\beta\theta\left(s-\frac{m_{1}^{2}}{\alpha}-\frac{m_{2}^{2}}{\beta}-\frac{m_{3}^{2}}{1-\alpha-\beta}\right)\frac{-\text{m3}}{64\pi^{4}(\alpha+\beta-1)^{2}}\biggl[-2\beta\text{m1}^{2}\left(3\alpha^{2}+\alpha(6\beta-7)+3\beta^{2}-7\beta+4\right)\nonumber \\
 & +\text{m1}\text{m2}(2\alpha+2\beta-3)(\alpha+\beta-1)+\alpha\biggl(\beta\left(2\text{m3}^{2}(3\alpha+3\beta-4)+3s(4\alpha+4\beta-5)(\alpha+\beta-1)\right)\nonumber \\
 & -2\text{m2}^{2}\left(3\alpha^{2}+\alpha(6\beta-7)+3\beta^{2}-7\beta+4\right)\biggr)\biggr]
\end{align}
}{\scriptsize \par}

\end{document}